\documentclass[aps,prb,showpacs,preprint]{revtex4}
\usepackage{graphicx}

\begin{document}
\title{ Statistics of electron tunneling in normal tunnel junctions}

 \author{P.J. Hakonen$^1$,  A. Paila$^1$, and E.B. Sonin$^{2}$}
  \affiliation{ $^1$Low Temperature Laboratory, Helsinki University of
Technology, FIN-02015 HUT, Finland \\
$^2$The Racah Institute of Physics, The Hebrew University of Jerusalem,
Jerusalem 91904, Israel}

\date{\today}

\begin{abstract}
Statistics of electron tunneling in normal tunnel junctions is
studied analytically and numerically taking into account circuit
(environment) effects. Full counting statistics, as well as full
statistics of voltage and phase have been found for arbitrary times
of observation. The theoretical analysis was based on the classical
master equation, whereas the numerical simulations employed standard
Monte-Carlo methods.
\end{abstract}
\maketitle

\section{Introduction}

Shot noise produced by electron current is known nearly 90 years
\cite{Shot}. The interest to this phenomenon was revived recently
due to great importance of shot noise for modern mesoscopic devices
\cite{BB}. Shot noise provides valuable information on the charge of
carriers, which are responsible for the electric current. Now
researchers are not content with the parametrization of noise using
variance or noise temperature and look for a complete picture of
charge noise given by all cumulants of the probability distribution
for the number of electrons, which traverse the junction during a
fixed period of time. The full set of moments or cumulants
determines {\em the full counting statistics}. The theoretical study
of the full counting statistics was initiated by Levitov and Lezovik
\cite{LL}, and recently these studies have been intensified and
brought about important results concerning the effect of environment
on the statistics (crossover from the voltage to the current bias)
\cite{KNB}. These studies focused on the long-time (low-frequency)
limit of the full counting statistics.

It is difficult to study full counting statistics experimentally,
but essential progress has been achieved in this direction too.
Experimentalists do not have direct access to counting statistics
itself, \textit{i.e.} they cannot determine those moments of time
when electrons cross the junction, but instead they can scan
voltage noise produced over a shunt by tunneling events. Until now,
only the first non-trivial cumulant, namely the third cumulant (skewness)
\cite{OddE,Rez}, has been detected experimentally. In the
literature, there has been discussion of other methods of noise
spectroscopy. It has been demonstrated theoretically and
experimentally that a Coulomb-blockade tunnel junction is an
effective probe of shot noise \cite{ours} (and of other types of
noise as well \cite {TNP}). On the other hand, Tobiska and Nazarov
\cite{TN} suggested to use a superconducting Josephson junction
close to the critical current as a threshold detector. This method
has  also been studied experimentally \cite{Pek}.

In tunnel junctions, shot noise is produced by discrete electrons,
which tunnel quantum-mechanically through a high potential barrier
and, thus, noise characterizes this quantum-mechanical process. This
was the reason to investigate shot noise with modern quantum-field
techniques \cite{KNB}. However, it is known that though quantum
mechanics is crucial for formulation of basic statistic properties
of electron transport, after the basic statistical properties of
electron tunneling have already been formulated, the following
statistical analysis, which should lead to knowledge of full
statistics can be done without any reference to quantum mechanics.
In particular, Nagaev \cite{Nag} have studied the effect of
environment (circuit) on the counting statistics using the classic
Boltzman-Langevin approach, which agrees with the results of the
quantum-mechanical analysis.

Our present work describes a further development of the classical
analysis of statistics of electron tunneling using the formalism of
the master equation and direct numerical simulations. The master
equation has been widely used for studying various problems of
statistics in physics and in other fields. The master equation for
the voltage probability distribution was used long time ago within
the framework of the ``orthodox theory'' of the Coulomb blockade
\cite{AL}. But it was mostly for the calculation of the dynamics of
single electron tunneling oscillations and the corresponding $IV$
curves at low bias, when Coulomb blockade modifies the $IV$ curve
essentially. The master equation has also been employed for studying
counting statistics in various mesoscopic setups \cite{BN}, but only
in the long-time (low-frequency) limit. Here our intention is to use
the master equation for full statistics of charge transfer through a
normal junction for any time scale. We write here {\em full
statistics} instead of {\em full counting statistics} since we have
analyzed not only the statistics of electron tunneling events
(counting) but also the statistics of voltage and phase
generated by these events. This is important, since, as mentioned
above, typical experiments probe voltage or phase instead of the
number of electrons. In the long-time (low frequency) limit, our
results completely agree with the previous quantum-mechanical
analysis \cite{KNB}. However, our final expressions valid at $T=0$
are not restricted to long times and, therefore, they provide
the full statistics for all time scales.

\section{Electric circuit and $IV$ curve} \label{elcirc}

\begin{figure}
  \begin{center}
    \leavevmode
    \includegraphics[width=0.7\linewidth]{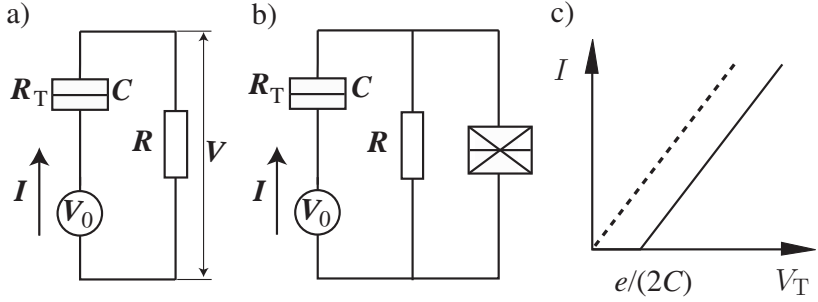}
    \caption{Electric circuit. a) Normal tunnel junction voltage-biased
    through the shunt resistance $R$. b) Parallel to the
shunt there is a Josephson junction, which probes the phase
difference at the shunt.  c) The $IV$ curve of the normal junction
in the limit of ideal voltage bias.}
  \label{fig1}
  \end{center}
  \end{figure}

We consider the simplest circuit, which is standard for studies of
shot noise in tunnel junctions: a tunnel junction of resistance
$R_T$ and of capacitance $C$ biased from an ideal voltage source
$V_0$ via a resistance $R$ (called shunt) in series (Fig.
\ref{fig1}a). The voltage across the junction is $V_T=V_0-V$ where
$V=IR$ is the voltage over the shunt resistor and $I=V_0/(R_T+R)$ is
the current through the junction. A basic
parameter for our analysis is the ratio $\alpha=R/R_T$. If $\alpha
\ll 1$ the circuit corresponds to an ideal voltage bias $V_0 \approx
V_T$, whereas if $\alpha \gg 1$ this is the case of ideal current
bias.

The standard method of noise spectroscopy is to measure voltage $V$
at the shunt \cite{OddE,Rez}. On the other hand, a Josephson
junction added parallel to the shunt (Fig. \ref{fig1}b) may probe
the phase difference fluctuations at the shunt \cite{ours,TNP}. We
assume that the both resistances, $R_T$ and $R$, clearly exceed the
quantum resistance $R_K=h/e^2$. According to the orthodox theory
\cite{AL} at $T=0$, in this limit the average rate of tunneling
through the junction  at $V_T>e/2C$ is $(V_T-e/2C)/eR_T$. The $ IV$
curve for an ideal voltage bias ($V_T\approx V_0$) is shown in Fig.
\ref{fig1}c. The junction is Coulomb blockaded as far as the voltage
$V_T$ does not exceed the Coulomb voltage offset $e/2C$.

\section{Full statistics for ideal voltage bias}
\label{IVB}

There is a straightforward procedure to find the full
statistics in the limit of ideal voltage bias exactly
\cite{ours,TNP}. Because of the constant voltage at the junction the
probability of tunneling is also constant and the full counting
statistics is given by the Poissonian distribution. The generating
function, which yields all moments and
cumulants, is
\begin{eqnarray}
F_c(\lambda,t)=\sum_n e^{\lambda
n}P_n(t)=e^{(V_Tt/eR_T)(e^\lambda-1)}=e^{(It/e)(e^\lambda-1)}~,
     \label{CS}     \end{eqnarray}
where $ P_n(t)=e^{-\bar n}\bar n^ n/n!$ is the Poissonian
distribution, $n$ is the number of tunneling events during the time
interval $t$. The average number  of events is determined by the
current: $\bar n=It/e$.

Though in the limit of ideal voltage bias the voltage at the junction $V_T \approx
V_0$ practically does not vary, the small voltage $V\ll V_0$ at the shunt
strongly fluctuates. In order to find the statistics of voltage $V$ let us
consider a sequence of $N$ tunneling events at random moments of time $t_j$
($j=1,2,...N$), which may occur during the long time interval $T$. Time $T$ is
connected with $N$ via the relation $N=IT/e$. Any tunneling produces the
voltage pulse at the shunt, so the voltage varies in  time
as
\begin{eqnarray}
V(t)=\sum_{j=1}^N {e\over C}e^{-(t-t_j)/\tau}\Theta(t-t_j)~,
       \label{volt}   \end{eqnarray}
where $\tau=RC$ and $\Theta(t)$ is the step function. Then one can find the
generating function for the voltage distribution averaging over
random moments of time $t_j$:
\begin{eqnarray}
F_v(\nu)=\left\langle e^{\nu v} \right \rangle =\left\langle e^{\nu \sum_j
e^{-(t-t_j)/\tau}} \right
\rangle=\prod_{j=1}^N \int_0^T {dt_j \over T} e^{\nu
e^{-(t-t_j)/\tau}\Theta(t-t_j)} \nonumber \\
=\left(1+{\Phi_v(\nu)\over T}\right)^N
=e^{\Phi(\nu)I/e}~,
    \label{VS}     \end{eqnarray}
where the dimensionless voltage $v=VC/e$ was introduced,
\begin{eqnarray}
\Phi_v(\nu)=\int_0^\infty dt \left(e^{\nu e^{-t/\tau}}-1\right)  = \tau [ \mbox{E}_1(-\nu ) -\gamma -\ln(-\nu )]
          \end{eqnarray} is the contribution of a single tunneling event, $\gamma$ is Euler's constant, and
$E_1(z)=\int_z^\infty(e^z-1)dz/z$ is the exponential integral.

Let us consider the random phase variation, which follows from Eq.
(\ref{volt}) and the Josephson relation
$d\varphi/dt=eV/\hbar$:
\begin{eqnarray}
\varphi(t)=\sum_{j=1}^N r\left[1-e^{-(t-t_j)/\tau}\right]\Theta(t-t_j)~,
          \end{eqnarray}
where $r=2\pi R/R_K$.  Our goal is to find the full statistics for the phase
difference $\Delta
\varphi(t)=\varphi(t+t_0)-\varphi(t_0)=\sum _j \delta \varphi_j$
during a fixed time interval $t$, where the contribution from the
tunneling at the moment $t_j$ is
\begin{eqnarray}
\delta \varphi_j=r\left[1-e^{-(t+t_0-t_j)/\tau}\right]\Theta(t+t_0-t_j)
-r\left[1-e^{-(t_0-t_j)/\tau}\right]\Theta(t_0-t_j)~.
          \end{eqnarray}
 Similar to the voltage statistics one should find the
generating function for the phase difference, which does not depend on $t_0$ after
averaging:
\begin{eqnarray}
F_\varphi(\xi,t)=\left\langle e^{\xi \Delta \phi(t)} \right
\rangle=\prod_{j=1}^N \int_0^T {dt_j \over T} e^{\xi \delta \varphi_j/r}
=\left(1+{\Phi_\varphi(\xi)\over T}\right)^N =e^{\Phi_\varphi(\xi,t)I/e}~,
    \label{PSv}      \end{eqnarray}
where $\phi=\varphi /r$ is the rescaled phase and
\begin{eqnarray}
\Phi_\varphi(\xi,t)= -t- \tau\left\{e^{\xi } \left[E_1(\xi ) - E_1 \left(\xi  e^{-t/\tau} \right)\right]  +E_1
\left[-\xi \left(1- e^{-t/\tau}
\right) \right] \right. \nonumber \\ \left.
+\gamma+\ln \left[-\xi \left(1- e^{-t/\tau}
\right) \right] \right\}~.
              \end{eqnarray}
We have introduced
the rescaled phase since after this, statistics of phase at long times is
identical to counting statistics (see below). Note that at $\xi= ir$ the
generating function
$F_\varphi(\xi,t)$ yields the phase correlator
$\langle e^{i\varphi(t)-i\varphi(0)}\rangle $, which determines the current through the normal tunnel junction in the $P(E)$
theory, whereas at $\xi=2 ir$ this yields the phase correlator for the Cooper pairs in the Josephson junction \cite{ours,TNP}.
At long times $t\gg \tau$ the phase statistics becomes identical to the
counting statistics (apart from the scaling factor $r$):
\begin{eqnarray} F_\varphi (\xi,t)=e^{(It/e)(e^{\xi }-1)}~.
      \label{PS}    \end{eqnarray} This is equal to $F_c (\lambda,t)$, Eq.
(\ref{CS}), with $\lambda= \xi $. On the other hand at short times $t \ll \tau$
the voltage does not vary essentially during the time $t$, and the phase
statistics should be identical to the voltage statistics. Indeed, at $t \ll
\tau$ Eq. (\ref{PSv}) yields the voltage generating function $F_v(\nu)$,  Eq.
(\ref{VS}),  with $\nu=\xi t /\tau $.

\section{Full voltage statistics from master equation}

In order to find the full voltage statistics for an arbitrary bias
we use the master equation for the voltage probability density $P(V,t)$:
\begin{eqnarray}
{\partial P(V,t)\over \partial t}-{1\over RC} {\partial \over \partial
V}[VP(V,t)] \nonumber \\ +{1\over eR_T}\left[\left(V_T-{e\over
2C}\right)P(V,t)-\left(V_T+{e\over 2C}\right)P\left(V-{e\over
C},t\right)\right]=0~.
            \end{eqnarray}
This equation directly follows from the master equation in Ref. \onlinecite{AL}
in the limit $T \to 0$ when tunneling can occur only in one direction. Our analysis
addresses the case of high currents when the voltage $V_T$ at the junction
always exceeds the Coulomb gap $e/2C$. We remind
that we consider the high-impedance circuit when the resistances $R$ and $R_T$
exceed the quantum resistance.

The equation  for the generating function ($v=VC/e$),
\begin{eqnarray} F_v(\nu)=\int e^{\nu v}P(V)dV~,
      \label{gf}    \end{eqnarray}
is obtained by integration of the master equation over the whole interval of
relevant voltages:
\begin{eqnarray}
{\partial F_v(\nu)\over \partial t}+{\nu\over RC}{\partial
F_v(\nu)\over \partial \nu}+{e^{\nu}-1\over eR_T}\left[{e\over C} {\partial
F_v(\nu)\over \partial \nu}-\left(V_0-{e\over 2C}\right)F_v(\nu) \right]=0~.
            \end{eqnarray}
In terms of the
dimensionless time $\tilde t=t/RC$  and the dimensionless voltage $v_0=VC/e$, the
equation becomes:
\begin{eqnarray} {\partial F_v(\nu)\over \partial \tilde t}+[\nu+\alpha(e^{\nu}-1)]{\partial F_v(\nu)\over \partial \nu}-\alpha
(e^{\nu}-1)\left(v_0-{1\over 2}\right)F_v(\nu)=0~.
            \end{eqnarray}
  In the stationary case this is an ordinary differential equation with respect
to $ \nu$, which has the following solution:
  \begin{eqnarray}
F_{v}( \nu)=\exp\left[\left(v_0-{1\over 2}\right)\int_0^{\nu}\frac{\alpha ( e^{z
}-1)}{z+\alpha  ( e^{z }-1)}dz
\right]~.
     \label{VStat}     \end{eqnarray} The solution satisfies the natural boundary condition that  $F_v(\nu)=1$ at $ \nu=0$.
According to Eq. (\ref{gf}) this provides a proper normalization of the probability density.

The limit $\alpha=R/R_K \to 0$  corresponds to the ideal-bias case
when the generating functions given by  Eqs.  (\ref{VS}) and
(\ref{VStat})  are identical keeping in mind that the current is
$I=(V_0-e/2C)/(R_T+R)$. In the general case of arbitrary $\alpha$,
the first three cumulants for the voltage at the shunt are:
\begin{eqnarray}
\langle \langle V \rangle \rangle ={e\over C}{d\ln F_{v}( \nu)\over d\nu}=IR~,
      \nonumber \\
 \langle \langle V^2 \rangle \rangle ={e^2\over C^2}{d^2\ln F_{v}( \nu)\over d\nu^2}={IR\over 2(1+\alpha)}{e\over C}~,
      \nonumber \\
      \langle \langle V^3 \rangle \rangle ={e^3\over C^3}{d^3\ln F_{v}( \nu^3)\over d\nu^3}={IR(2-\alpha)\over
6(1+\alpha)^2}{e^2\over C^2}~.
     \label{VolCum}           \end{eqnarray} The first cumulant is simply the average voltage at the shunt. The third cumulant
(skewness) of the voltage distribution changes sign at $\alpha=1/2$.
The width of the distribution is determined by the smallest from the
two voltages: the shunt voltage $IR$ for the ideal voltage bias
($\alpha \to 0$), or the junction voltage $IR_T$  for the ideal
current bias ($\alpha \to \infty$).

\section{Full phase statistics} \label{FPS}

In order to find the full phase statistics one should consider the master
equation for voltage and phase. We introduce the probability
density $P(V,\varphi,t)$ for the shunt voltage $V$ and the corresponding phase
$\varphi(t)=(e/\hbar)\int^t V(t')dt'$. The master equation is:
\begin{eqnarray}
{\partial P(V,\varphi,t)\over \partial t}+{eV\over \hbar}{\partial
P(V,\varphi,t)\over \partial \varphi}-{1\over RC} {\partial \over \partial
V}[VP(V,\varphi,t)] \nonumber \\ +{1\over eR_T}\left[\left(V_T-{e\over
2C}\right)P(V,\varphi,t)-\left(V_T+{e\over 2C}\right)P\left(V-{e\over
C},\varphi,t\right)\right]=0~.
            \end{eqnarray}
The full statistics of voltage and phase is determined by the generating function
for the united phase+voltage probability distribution:
\begin{eqnarray}
 F_{v\varphi}(\xi,\nu,t)=\int e^{ \xi \phi +\nu v}P(V,\varphi,t)d\varphi\,dV~,
            \end{eqnarray}
where $\phi=\varphi /r$ is the rescaled phase and $v=VC/e$ is the rescaled
voltage. The master equation yields the following equation for the generating
function
\begin{eqnarray}
{\partial   F_{v\varphi}(\xi,\nu,\tilde t)\over \partial \tilde
t}+\left[\nu-\xi+\alpha (e^{\nu } -1)\right] {\partial
F_{v\varphi}(\xi,\nu,\tilde t)\over \partial \xi} -\alpha (e^{\nu }
-1)\left(v_0-{1\over 2}\right) F_{v\varphi}(\xi,\nu,\tilde t)=0~.
                            \end{eqnarray}
Here we use dimensionless time $\tilde
t=t/RC$. There is a well known analogy of the phase with a coordinate of
a diffusing particle and of the voltage with a particle velocity. Correspondingly,
the phase distribution can never be stationary and permanently expands. Thus
the generating function is always time-dependent. Performing the
Laplace transformation,
\begin{eqnarray}
 F_{v\varphi}(\xi,\nu,s)= \int_0^\infty e^{-s\tilde t}F_{v\varphi}(\xi,\nu,\tilde
t) d\tilde t~,
                            \end{eqnarray}
the equation for the generating function becomes
\begin{eqnarray} sF_{v\varphi}(\xi,\nu,s)+\left[\nu-\xi+\alpha (e^{\nu }
-1)\right] {d F_{v\varphi}(\xi,\nu,s)\over d\xi} -\alpha (e^{\nu }
-1)\left(v_0-{1\over 2}\right)
F_{v\varphi}(\xi,\nu,s) \nonumber \\
=F_{v\varphi}(\xi,\nu,t)|_{t=0}~,
               \label{p+v}             \end{eqnarray}
where $F_{v\varphi}(\xi,\nu,t)|_{t=0}$ is the initial value of the generating
function. The solution of this nonuniform differential equation is
 \begin{eqnarray} F_{v\varphi}(\xi,\nu,s) =F_0(\xi,\nu,s)\int_{\xi_1+\delta}^\nu {F_{v\varphi}(\xi,x,t)|_{t=0}\over
[-\xi+x+\alpha(  e^{x} -1) ] F_0(\xi,x,s)}dx~,
          \label{pvsol}                \end{eqnarray}
where $\xi_1$ is a zero of the denominator in the
integrand function,
 \begin{eqnarray} -\xi+\xi_1+\alpha(  e^{\xi_1} -1)=0~,
          \label{pole}                \end{eqnarray}
and
\begin{eqnarray} F_0(\xi,\nu,s)=\exp \left[\int_{\xi_1+\delta }^\nu {-s+\left(v_0-{1\over 2}\right) \alpha (e^{x } -1) \over
-\xi+x+\alpha (e^{x } -1)}  dx \right]
     \label{unif}                    \end{eqnarray}
is the solution of the uniform equation when the right-hand part of Eq.
(\ref{p+v})  vanishes. In order to cut divergence at the pole at $x \to \xi_1$, an
 infinitely small constant $\delta$ was introduced. In final expressions this divergence is canceled and the limit $\delta \to 0$
yields convergent results.  One can check that the solution Eq. (\ref{pvsol}) satisfies the boundary condition
$F_{v\varphi}(\xi,\nu,t)=1$ at $\xi=\nu=0$, which provides a proper normalization of the probability density. The Laplace
transform of this condition is $F_{v\varphi}(\xi,\nu,s)=1/s$ at $\xi=\nu=0$.    In order to obtain the full phase statistics from
the solution  Eq. (\ref{pvsol}) one should choose the initial condition that
\begin{eqnarray}
F_{v\varphi}(\xi,\nu,t)|_{t=0} = F_{v}( \nu)~,
                           \end{eqnarray}
where $F_{v}( \nu) $ is the generating function for the stationary shunt voltage
distribution given by Eq. (\ref{VStat}). Independence of this function on $\xi$ means that we fixed the phase $\varphi=0$ at the
initial moment of time $t=0$. Finally the full phase statistics after averaging over the voltage is given by:
 \begin{eqnarray}
F_{\varphi}(\xi,s) =F_{v\varphi}(\xi,0,s) =F_0(\xi,0,s)\int_{\xi_1+\delta}^0
{F_{v}( x)\over [-\xi+x+\alpha( e^{x} -1) ] F_0(\xi,x,s)}dx~.
        \label{phas}                 \end{eqnarray}
Taking derivatives with respect to the variable $\xi$ conjugate to the phase
difference one may obtain any moment or cumulant for the phase-difference
probability. But  even for the second or the third cumulant taking derivatives is
a tiresome procedure in general. So we restrict the further analysis with
some limiting cases.

\subsubsection*{Long-time limit}

This case was also analyzed quantum-mechanically \cite{KNB}. We shall see that
our general classical analysis completely agrees with it.

In the case of the long-time asymptotics the most important contribution to the integral in (\ref{phas}) comes from the close
vicinity of the pole determined by Eq. (\ref{pole}). The solution of the uniform equation, Eq. (\ref{unif}), can be also  reduced
to the contribution of the pole, and
\begin{eqnarray}
F_0(\xi,\nu,s)\approx \exp \left[\int_{\xi_1+\delta }^\nu {-s+s_0 \over
-\xi+x+\alpha (e^{x } -1)}  dx \right]
\approx \left( {\delta \over
\nu-\xi_1}\right)^ {(s_0 -s)/(1+\alpha)}~,
                          \end{eqnarray}
where
\begin{eqnarray} s_0= \left(v_0-{1\over 2}\right) \alpha (e^{\xi_1} -1)~.
                          \end{eqnarray}
For long-time asymptotics the initial condition for the generating function is not
essential, and one can assume for simplicity that $F_{v\varphi}(\xi,x,t)|_{t=0}=F_{v}( x) =1$  in Eq. (\ref{phas}). This means
that at $t=0$ the voltage and phase difference at the shunt are zero. Finally the Laplace transform of the generating function for
phase probability is approximated at long times with
\begin{eqnarray} F_{\varphi}(\xi,s) \approx {1 \over s-s_0}~.
                          \end{eqnarray}
After the inverse Laplace transformation the generating function in the time
presentation is
\begin{eqnarray} F_{\varphi}(\xi,t) \approx e^{s_0 \tilde t}~.
      \label{FSlt}                    \end{eqnarray}
This fully agrees with the result obtained by Kindermann {\em et al.}
\cite{KNB} from the quantum-mechanical analysis. This is a manifestation of a
simple law, which Kindermann {\em et al.} have formulated for the full statistics
of two devices connected in series with the ideal voltage source. If the full
statistics (either counting statistics or phase statistics) for any of two devices
connected directly to the voltage source without another device are
$F_1(\xi_1)$ or $F_2(\xi_2)$ respectively, then for the two devices connected
together the full statistics is $F(\xi)=F_1(\xi_1)=F_2(\xi_2)$
 at the condition that
$\xi=\xi_1+\xi_2$. In our case, the two devices are a normal tunnel junction and a
macroscopic resistor (shunt). Then $F_1(\xi_1)=e^{(It/e)(e^{\xi_1 }-1)}$ follows
from the phase statistics of the junction at the ideal voltage bias, Eq.
(\ref{PS}), whereas $F_2(\xi_2)=e^{(It)
\xi_2/e}$ corresponds to a macroscopic resistor (shunt) under fixed voltage bias
$IR$. Equation (\ref{pole}) is identical to the equation
$F_1(\xi_1)=F_2(\xi-\xi_1)$.

The generating function Eq. (\ref{FSlt}) yields the following first 4 cumulants:
 \begin{eqnarray}
\langle \langle \varphi \rangle \rangle =r {d\ln F_{\varphi} \over d\xi }
=r{ds_0 \over d\xi }\tilde t =r{ds_0 \over d\xi_1}{d\xi_1 \over d\xi}\tilde t
=r\left(v_0-{1\over 2}\right)\tilde t {\alpha  \over 1+ \alpha } =r{It\over e}~,
           \label{c1}              \end{eqnarray}
\begin{eqnarray}
\langle \langle \varphi^2 \rangle \rangle =r^2 {d^2\ln F_{\varphi} \over d\xi^2 }=r^2 \left(v_0-{1\over 2}\right)\tilde t {\alpha
\over (1+ \alpha)^3 } =r^2{It\over e} {1  \over (1+ \alpha)^2 }~,
          \label{c2}                \end{eqnarray}
\begin{eqnarray}
\langle \langle \varphi^3 \rangle \rangle =r^3 {d^3\ln F_{\varphi} \over d\xi^3
}=r^3
\left(v_0-{1\over 2}\right)\tilde t{\alpha (1-2\alpha)  \over (1+ \alpha)^5 }
=r^3{It\over e} {1-2\alpha  \over (1+ \alpha)^4 }~,
          \label{c3}                \end{eqnarray}
\begin{eqnarray}
\langle \langle \varphi^4 \rangle \rangle =r^4 {d^4\ln F_{\varphi} \over d\xi^4
}=r^4\left(v_0-{1\over 2}\right)\tilde t {\alpha (1-8\alpha+6\alpha^2)  \over (1+
\alpha)^7 } =r^4{It\over e}{ 1-8\alpha+6\alpha^2  \over (1+ \alpha)^6 }~.
            \label{c4}              \end{eqnarray}

Kindermann {\em et al.} \cite{KNB} have shown that in the limit of ideal current
bias ($\alpha \gg 1$)  the statistics of the phase at a contact of
arbitrary transparency corresponds to the Pascal distribution \cite{Spi}. If the
contact transparency is very low (the case of tunnel junction) the Pascal
distribution is reduced to  the chi-square distribution. One can check that this
fully  agrees with our analysis, and the found full phase statistics corresponds
to the chi-square distribution for phase probability.

         \subsubsection*{Short-time limit}

The full phase statistics for short time intervals directly follows
from a plausible assumption that the voltage does not vary during
the observation time. This means that the generating function
$F_\varphi(\xi)$ for the phase  is  equal to  the generating function
$F_v(\nu)$ for the voltage, Eq. (\ref{VStat}), with $\nu= \xi
t/\tau$ as directly proved for an ideal voltage bias in Sec.
\ref{IVB}.

Our analysis shows that the crossover between the long-time and the short-time
behavior is governed by the relaxation time $\tilde \tau=CRR_T/(R_T+R)$. This time
is different from the relaxation time $\tau=RC$, which characterizes the
electron transport in the circuit between tunneling events. The two relaxation
times coincide only in the ideal-voltage-bias limit.

 \section{Counting statistics} \label{fCS}

In order to find the full counting statistics we introduce the density $P_n(V,t)$
of probability that during the time interval $t$
$n$ electrons tunneled through the junction and that in the end of the interval
the voltage at the shunt is $V$.  The master equation for this probability is
\begin{eqnarray}
{\partial P_n ( V,t)\over \partial t}-{1\over RC} {d\over
dV}[VP_n(V,t)]  \nonumber \\
+{1\over eR_T}\left[\left(V_T-{e\over
2C}\right)P_n(V,t)-\left(V_T+{e\over 2C}\right)P_{n-1}\left(V-{e\over
C},t\right)\right]=0~.
            \end{eqnarray}
Let us introduce the generating function:
\begin{eqnarray}
F_{cv}(\lambda,\nu,t)=\sum_n e^{\lambda n} \int e^{\nu  v}P_n
( V,t) dV~.
            \end{eqnarray}
The Laplace transform of the equation for the
generating function is (the dimensionless time variable
$\tilde  t=t/RC$ was used):
 \begin{eqnarray}
s F_{cv}(\lambda,\nu,s)+{dF_{cv}(\lambda,\nu,s)\over d\nu}
\left[\nu + \alpha (e^{\lambda+\nu } -1)\right]-v_0\alpha (e^{\lambda+\nu }-1
)F_{cv}(\lambda,\nu,s)\nonumber \\
=F_{cv}(\lambda,\nu,t)|_{t=0}~.
                            \end{eqnarray}
The left-hand side of this equation is
identical to the left-hand side of Eq. (\ref{p+v}) after transformation
$\xi=\lambda+\nu$. The full counting statistics corresponds to the limit $\nu \to
0$. Eventually the generating function can be easily obtained from Eq.
(\ref{phas}) for the full phase statistics:
 \begin{eqnarray}
F_c(\lambda,s) =F_0(\lambda,0,s)\int_{\xi_1+\delta}^0
{F_{v\varphi}(\lambda+x,x,t)|_{t=0}\over [-\lambda+x+\alpha(  e^{x} -1) ]
F_0(\lambda,x,s)}dx~.
        \label{FCS}                 \end{eqnarray}

\subsubsection*{Long times}

The only difference between the phase statistics and the full counting
statistics appears in the initial boundary condition. Since for the
long-time asymptotic the initial boundary condition is not
essential, in this limit the cumulants  for the full counting
statistics can be  obtained from those for the full phase statistics
[Eqs. (\ref{c1}-\ref{c2})] by simple rescaling. Namely, the relation
between two types of cumulants is $\langle \langle n^k \rangle
\rangle=\langle \langle \phi^k \rangle \rangle =\langle \langle \varphi^k
\rangle \rangle/r^k $. This relation was derived by Kindermann {et al.}
\cite{KNB}, but it was approximate in their
quantum-mechanical analysis. In our purely classical approach this relation is
exact.
\subsubsection*{Short times}

In analogy with the short-time limit ($t\ll \tilde \tau$) of the phase statistics we use the fact that the voltage does not vary
essentially  during the time $t$. If one considers a subensemble  of
realizations, which correspond to some fixed voltage $V_T$ at the junction during
the time interval $t$, the distribution of numbers of tunneling events is
Poissonian and the generating function is given by Eq.  (\ref{CS}). But one should
take into account that the voltage $V_T$ at the junction fluctuates.
Thus one should average over
$V_T=V_0-V$, and the generating function is given by
\begin{eqnarray}
F_c(\lambda,t)=\int e^{(V_Tt/eR_T)(e^\lambda-1)}P(V_T)dV_T   =\int
e^{[(V_0-V)t/eR_T(e^\lambda-1)}P(V)dV~.
       \end{eqnarray}
One can see that the generating function for the counting statistics is directly
connected with the generating function for the voltage, Eq. (\ref{gf}):
\begin{eqnarray}
F_c(\lambda,t)=e^{(V_0)t/eR_T(e^\lambda-1)}F_v(\tilde \nu)~,
       \end{eqnarray}
where
\begin{eqnarray}
\tilde \nu=-{t\over R_TC}(e^\lambda -1)~.
       \end{eqnarray}
This allows to obtain expressions for any cumulant of the full counting statistics
using the expressions for voltage cumulants, Eq.
(\ref{VolCum}).
The first three of them are:
\begin{eqnarray}
\langle \langle n \rangle \rangle ={d \ln F_c\over d\lambda}=
{V_0t\over eR_T}+ {d \ln F_v\over d\tilde \nu}{d\tilde \nu\over
d\lambda}\\ = {V_0t\over eR_T}-\langle \langle V \rangle \rangle
{C\over e}{t\over R_T C}={It\over e}~,
          \end{eqnarray}
\begin{eqnarray}
\langle \langle n^2 \rangle \rangle ={d^2 \ln F_c\over d\lambda^2}= {V_0t\over
eR_T}+ {d \ln F_v\over d\tilde \nu}{d^2\tilde
\nu\over d\lambda^2}+ {d^2 \ln F_v\over d\tilde \nu^2}\left({d\tilde \nu\over
d\lambda}\right)^2={V_0t\over eR_T}-\langle \langle V \rangle \rangle {C\over
e}{t\over R_T C}\nonumber \\ +\langle \langle V^2 \rangle \rangle {C^2\over
e^2}\left({t\over R_T C}\right)^2
 ={It\over e}\left[1+{t\over \tilde \tau}{\alpha^2\over
2(1+\alpha)^2} \right]~,
         \end{eqnarray}
\begin{eqnarray}
\langle \langle n^3 \rangle \rangle ={d^3 \ln F_c\over d\lambda^3}= {V_0t\over
eR_T}e^\lambda+ {d \ln F_v\over d\tilde
\nu}{d^3\tilde \nu\over d\lambda^3}+3 {d^2 \ln F_v\over d\tilde \nu^2}{d^2\tilde
\nu\over d\lambda^2} {d\tilde \nu\over d\lambda}+ {d^3 \ln F_v\over d\tilde
\nu^3}\left({d\tilde \nu\over d\lambda}\right)^3 \to {It\over e} \nonumber \\ +3
\langle \langle V^2
\rangle \rangle {C^2\over e^2}\left({t\over R_T C}\right)^2 -\langle \langle V^3
\rangle \rangle {C^3\over e^3}\left({t\over R_T C}\right)^3\nonumber \\
={It\over
e}\left[1+{t\over \tilde \tau}{3\alpha^2\over 2(1+\alpha)^2}-\left({t\over \tilde
\tau }\right)^2 {\alpha^3(2-\alpha)\over 6(1+\alpha)^4} \right]~ .
            \end{eqnarray}
These expressions demonstrate that at short times $t\ll \tilde
\tau=CRR_T/(R_T+R)$ the counting statistics becomes Poissonian even in the limit
of current bias $\alpha \to \infty$. Thus for short times the circuit
(environment) effects are not so essential as for long times.
It is worthwhile to note
that ``short'' times $t\ll \tilde \tau$ in reality are not necessarily short
compared with the average time $e/I$ between tunneling events since we consider
the case of high currents $I$, and very many electrons may tunnel during the
``short'' time $t\ll \tilde \tau$.

\section{Numerical simulation}

In our computational model, the time dependence of charge $Q(t)=CV_T(t)$
on the junction with capacitance $C$ is obtained by integrating the equation
\begin{equation}
\frac{dQ}{dt}=\frac{V_{0}-Q/C}{R}-\left(
\frac{dQ}{dt}\right) _{T}, \label{charge}
             \end{equation}

\noindent where the first term on the right represents charge
relaxation through the shunt resistor $R$, and the latter term represents
tunneling current in the tunnel junction. According to Sec. \ref{elcirc} at
$V_T>e/2C$ the average tunneling rate
$\langle dQ_I/dt _{T}\rangle$ is  $(Q/e-1/2)/RC$. Employing standard
Monte-Carlo methods, this average rate is used to generate tunneling events,
which are considered to take place instantaneously on the time
scales of other processes. A ready-made, Mersenne-Twisters-type
random number generator \cite{Mersenne} was employed in our Fortran
code.

The simulation was performed using parameter values close to
standard mesoscopic tunnel junctions, namely $R_T=10$ k$\Omega$ and
$C=1$ fF which corresponds to a Coulomb voltage of $e/2C \sim 0.1$
mV. We used $V_0=5$ mV for the bias voltage, while the length of the
time record was typically set to 1 ns. With these values our time
trace displayed altogether a few hundred tunneling events. The
relaxation rate $\tilde \tau=CRR_T/(R_T+R) $ normally does not
exceed 10 ps, and our simulation is basically in the long time limit
discussed in Secs. \ref{FPS} and \ref{fCS}. The time step in the
integration was $\Delta t =10^{-14} - 10^{-13}$ s. The simulation
was initialized for 50000 iterations before starting the calculation
of the actual time traces. For making distributions, the calculation
was repeated for $5-10 \cdot 10^5$ times using the previous
simulation as an initialization for the next one.

Fig. 2 characterizes the skewness of the distribution of the voltage
$V$ over the shunt resistor obtained in our simulations. The inset
displays the logarithm of a voltage distribution $\ln P(V)$,
calculated at $\alpha=0.5$. The asymmetry is clearly visible in the
wings of the $\ln P(V)$-curve. The solid curve depicts the
theoretical curve for $\langle \langle V^3 \rangle \rangle$ obtained
from Eq. (\ref{VolCum}).
 The master equation
approach is seen to agree with the Monte-Carlo simulation within the
scatter of the data points.


Phase $\varphi(t)$ was calculated from the simulated voltage trace
by numerically evaluating $\int dt V(t)$ using a 3-point Simpson
rule. Fig. 3 displays the result for the ratio $\langle
\langle\varphi^3\rangle \rangle /\langle \langle\varphi^2\rangle
\rangle$; we have chosen this ratio as both the nominator and the
denominator are proportional to $I$, eliminating the bias dependence
off from it. The behavior at small values of $\alpha$ is illustrated
in the inset in more detail. The simulated data are seen to display
a change of sign in a similar manner as in the theory at
$\alpha=0.5$. Notice that the functional dependence here coincides
with the current fluctuation results in the low-frequency limit
\cite{KNB}.

Fig. 4 illustrates the ratio of the skewness to the variance for the counting
statistics as a function of the parameter $\alpha$ in the long-time
limit,
\emph{i.e.} when the length of the time trace exceeds the relaxation time $\tilde
\tau=CRR_T/(R_T+R) $. In general, we find a good agreement between our
Monte-Carlo simulation and the ratio calculated from Eqs.
(\ref{c2}) and (\ref{c3}) with help of the relation $\langle \langle n^k \rangle
\rangle =\langle \langle \varphi^k
\rangle \rangle/r^k $, except for a small offset at
$\alpha > 2$. The inset magnifies the results at $\alpha < 1.5$: the data on
$\langle \langle n^3\rangle
\rangle /\langle \langle n^2\rangle \rangle $ is seen to approach the Poisson
result $\langle \langle n^3\rangle \rangle /\langle \langle n^2\rangle \rangle
=1$ as expected when $\alpha \rightarrow 0$.

\begin{figure}
  \begin{center}
    \leavevmode
    \includegraphics[width=0.5\linewidth]{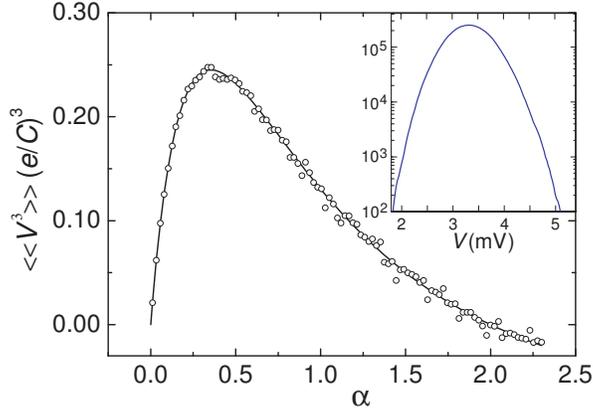}
    \caption{Skewness of the voltage distribution over the shunt resistor as a
    function of $\alpha=R/R_T$. The solid curve is calculated using Eq.
(\ref{VolCum}).
    The inset displays the probability distribution $P(V)$ for voltage at $\alpha=0.5$ in arbitrary units.}
  \label{fig2}
  \end{center}
  \end{figure}

\begin{figure}
  \begin{center}
    \leavevmode
    \includegraphics[width=0.5\linewidth]{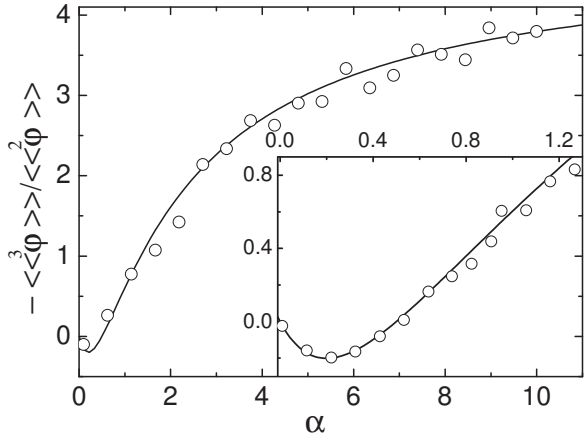}
    \caption{Illustration of phase statistics over the shunt resistor in terms of the ratio
$\langle \langle \varphi^3\rangle \rangle /\langle \langle \varphi^2\rangle \rangle $. The solid curve is calculated using
Eqs. (\ref{c2}) and (\ref{c3}).  The inset displays a magnification of the initial
part of the data.}
  \label{fig3}
  \end{center}
  \end{figure}

\begin{figure}
  \begin{center}
    \leavevmode
    \includegraphics[width=0.5\linewidth]{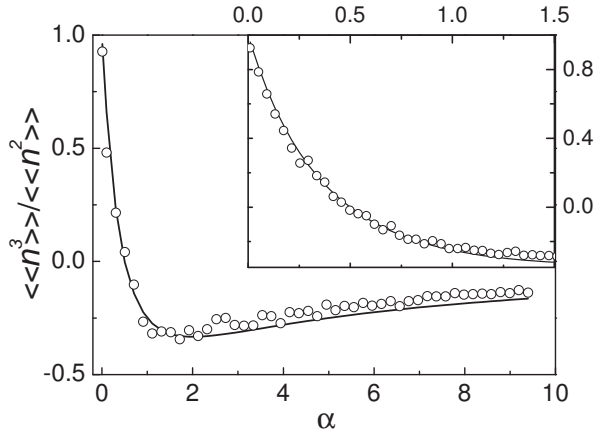}
    \caption{Ratio of the second and the third cumulants of the counting
    statistics distribution. The solid curve is calculated using
Eqs. (\ref{c2}) and (\ref{c3}) and the relation $\langle \langle n^k \rangle
\rangle =\langle \langle \varphi^k
\rangle \rangle/r^k $.  The inset displays a
magnification of the initial part of the data.}
  \label{fig4}
  \end{center}
  \end{figure}

\section{Conclusions}

Our classical approach based on the master equation provides the
full statistics of electron transport through a normal junction in a
high-impedance circuit. In addition to the full counting statistics,
the statistics of voltage and phase at the shunt resistor was found.
The analysis is valid for any time interval of observation. The
results are in full agreement with the results of the
quantum-mechanical analysis performed in the long-time limit
\cite{KNB}. In particular, the crossover of full counting statistics
from the Poissonian in the voltage-biased limit to the chi-square
distribution in the current-biased case was obtained. The identity
of counting statistics and phase statistics, which was found as an
approximate result of the quantum-mechanical analysis \cite{KNB},
was proven to be exact within classical approach. Our analysis shows
that strong environment (circuit) effects on counting statistics
become much weaker at short time scales (high frequencies). We have
performed numerical simulations using Monte-Carlo methods, which
fully agree with our analytical results.

\section*{Acknowledgements}

We acknowledge fruitful discussions with T. Heikkil\"a, F. Hekking,
and J. Pekola. This work was supported by the Academy of Finland and
by the Large Scale Installation programme ULTI-IV of the European
Union.

\end{document}